\documentclass[12pt]{iopart}

\usepackage[authoryear]{natbib}
\usepackage{graphicx}
\usepackage{url}
\usepackage{comment}

\begin{document}

\title[Projections of standardised energy indices]{Projections of standardised energy indices in future climate scenarios}

\author{E Dolores-Tesillos$^1$, N Otero$^2$ and S Allen$^3$}

\address{$^1$ Institute of Geography, Oeschger Centre for Climate Change Research, University of Bern, Hallerstrasse 12, Bern, Switzerland}
\address{$^2$ Fraunhofer Heinrich-Hertz-Institute – Fraunhofer HHI, Berlin, Germany}
\address{$^3$ Seminar for Statistics, ETH Zurich, Zurich, Switzerland}
\ead{edgar.dolores@unibe.ch}
\vspace{10pt}
\begin{indented}
\item[]September 2024
\end{indented}

\begin{abstract}
    Renewable energy is becoming an increasingly important component of energy systems. However, renewable energy production is heavily dependent on the prevailing weather conditions, which are changing as a result of climate change. It is therefore necessary to build energy systems that are robust to energy shortages caused by weather-dependent changes to energy demand and renewable energy production. To design such systems, we must monitor how changes in the climate are expected to influence future energy production and demand; this is important for policymakers to decide when, where, and by how much renewable energy installed capacities should be increased, for example. In this paper, we study the behaviour of standardised energy indices in future European climate projections, and use this to monitor how characteristics of energy production droughts in Europe are expected to change in the future. We use these results to make suggestions regarding how the energy mix should be adapted in the future to decrease the risk of energy production droughts.
\end{abstract}

%
\vspace{2pc}

\section{Introduction}

A significant decarbonisation of the energy system is crucial to mitigate impacts associated with climate change \citep{Bruckner2014}. The EU and its Member States are moving towards a climate-neutral future, with a commitment that renewable energy will comprise at least 42.5\% of national energy mixes by 2030 \citep{EEA2023}. With this transition towards renewable energy in progress, it is essential that we increase the flexibility of the EU power system to ensure a reliable energy supply, minimising the risk of power outages \citep{Cronin2018}. In particular, it is critical to balance renewable energy generation and energy consumption \citep{vonBremen2010}.

Sources of renewable energy are sensitive to fluctuations in the weather and climate, adding a layer of complexity to this balancing act. Solar and wind energy, for example, are the fastest-growing electricity sources \citep{Cronin2018,EEA2023}, and are expected to play a yet more significant role in Europe's future renewable energy systems. However, both are governed by weather conditions with high spatial and temporal variability, giving these sources a variable nature that poses a challenge for consistent energy production  \citep{vonBremen2010,RaynaudEtAl2018}.

A number of studies have investigated how the prevailing weather conditions affect the production of renewable energy \citep[e.g.][]{Troccoli2010, Thornton2017, Grams2017, RaynaudEtAl2018, Bloomfield2020}. These studies highlight the vulnerability of systems to weather conditions that result in abnormally low energy production and/or high electricity demand, which may in turn lead to energy shortages, or \textit{energy droughts}. Energy droughts have recently received considerable attention in the literature due to their relevance to power systems and energy markets \citep{RaynaudEtAl2018, Jurasz2021, Otero2022, BrackenEtAl2023}. It is common to consider energy production droughts, which correspond to when renewable energy production is abnormally low, and energy supply droughts, when the energy demand is abnormally high relative to production \citep{RaynaudEtAl2018}. 

Due to the weather-dependence of energy demand and renewable energy production, changes in the climate are expected to affect the characteristics of energy droughts. Recent studies have therefore investigated how renewable energy production will change under future climate conditions \citep[e.g.][]{Yang2022}, and how this will influence the frequency and duration of energy production droughts \citep{KapicaEtAl2024, ZuoEtAl2024}. In these studies, energy production droughts are defined as instances where the renewable energy production falls below a pre-determined threshold, corresponding to a quantile of previously observed values. However, the scale of renewable energy production (and energy demand) depends on several factors, such as the region of interest; a daily production of 500 GWh might be abnormally low in one region, but abnormally high in another, depending on the region's climate and installed capacities. This makes it challenging to compare the characteristics of energy droughts across different regions. For example, how does one systematically assess whether climate change will have a more profound effect on the severity of energy droughts in one region compared to another, if the interpretation of the drought's severity depends on the region?

\cite{AllenOtero2023} therefore suggest defining energy droughts in terms of standardised energy indices \citep[see also][]{StoopEtAl2024}. The standardised indices project renewable energy production to a standardised scale, facilitating comparisons between regions with different climates and installed capacities. In this paper, we investigate how the behaviour of standardised energy indices, and the resulting energy droughts, evolve in Europe as a result of climate change. Results are presented for one climate change scenario and several General Circulation Models using CMIP5 data (see Section \ref{sec:method}). We obtain similar results to previous studies regarding the frequency and duration of future European energy droughts, but by employing standardised indices, we can additionally make conclusions about the behaviour of standardised energy indices and the magnitude of the corresponding droughts. We further extend previous studies by investigating how installed capacities could be adapted so that energy production droughts do not become more common or severe in the future.

Due to the available data, we restrict attention to energy production droughts, and do not consider energy supply droughts. As such, the term ``energy droughts'' henceforth corresponds to energy production droughts. The definition of energy droughts in terms of standardised indices is presented in Section \ref{sec:sei}. Section \ref{sec:method} presents the data and methods used to study future energy drought characteristics in this work, the results of which are presented in Section \ref{sec:res}. These results are summarised in Section \ref{sec:conc}, before Section \ref{sec:disc} discusses the implications of these findings for energy policy and planning, and calls for further work to make more energy data publicly available. Additional results can be found in the supplementary material.

\section{Energy drought definitions}\label{sec:sei}

Following \cite{AllenOtero2023}, we define energy production droughts in terms of the Standardised Renewable Energy Production Index (SREPI). The SREPI is defined as

\begin{equation*}
	\mathrm{SREPI}(P_{t}) = \Phi^{-1} \left( F_{P}(P_{t}) \right),
\end{equation*}

where $P_{t}$ denotes the renewable energy production at time $t$, $F_{P}$ is an estimate of the cumulative distribution function of renewable energy production, and $\Phi^{-1}$ is the quantile function of the standard normal distribution. Throughout, we take $F_{P}$ to be the empirical distribution function corresponding to a time series of past observations, calculated separately for each season. 

Energy production droughts are defined as instances where the SREPI falls below -1.28, which occurs when renewable energy production is lower than 90\% of previously observed production values. 
Similar definitions have become standard when defining hydro-meteorological droughts \citep[see e.g.][]{Mckee1993,Vicente2010}. Lower thresholds can also be chosen to define more severe droughts, results for which are presented in the supplementary material.

In this study, the past observations used to estimate $F_{P}$ are daily time series of renewable energy production prior to the current day. This means that the standardised energy indices, and hence the corresponding energy droughts, are defined in terms of the current climate. This is useful when studying properties of energy droughts in the future, since it allows us to investigate how the characteristics of energy droughts will evolve if we maintain the current energy system; we can then deliberately study how specific changes to the energy system, such as increasing installed renewable energy capacities, affect these characteristics in terms of today's energy system.
 
\cite{KapicaEtAl2024} and \cite{ZuoEtAl2024} alternatively study the behaviour of energy droughts in climate change scenarios when droughts are defined as the non-exceedance of a fixed production threshold. If this threshold is defined in terms of quantiles of previous observations, as is commonly the case, then this is equivalent to defining energy droughts using standardised energy indices. The advantage of standardised indices is that the indices are defined on a standardised scale, meaning they can easily be compared for different locations and under different climatic conditions; see \cite{AllenOtero2023} and \cite{KittelSchill2024} for more thorough discussions on different ways to define energy droughts.

Using the definition above, each time step is or is not in an energy drought state. The \textit{duration} of the drought can be defined as the number of consecutive time steps in a drought state. Similarly, the \textit{magnitude} of the drought can be quantified by the cumulative SREPI value recorded during the drought \citep{AllenOtero2023}; this assumes that the most impactful droughts are those that persist for a long time, even if relatively weak at individual time steps, or when the production is severely lower than expected, even if only for a short time (or both). The standardised index on a given day quantifies the \textit{intensity} of the drought, and we also consider the average index value over the drought event as a measure of the average drought intensity \citep{BrackenEtAl2023}. We additionally employ the Compound Drought Magnitude proposed by \cite{BrackenEtAl2023}, for which we calculate the SREPI separately for wind and solar production, and then average these two indices during drought events; lower values therefore correspond to droughts for which the wind and solar production are both abnormally low. The SREPI therefore provides a convenient framework with which to study the frequency, duration, magnitude, and intensity of drought events. In the following, the SREPI is calculated using the \texttt{SEI} package in \texttt{R} \citep{AllenOtero2023b}.

\section{Methods}\label{sec:method}

To study future energy droughts (EDs), we consider daily data on solar photovoltaic power generation (SPV) and onshore wind power generation (WON) in capacity factor units. Offshore wind generation is not considered due to data availability. The data has been extracted at a sub-country level from the \cite{cs32021} \citep[see also][]{dubus2023energy}, with ``sub-country level'' referring to level 2 of the European Union's NUTS (Nomenclature des Unit{\'e}s Territoriales Statistiques) classification, which corresponds to geographical regions for regional policies.

Following \cite{KapicaEtAl2024}, we refer to data between 1970 and 2020 as the \textit{present day}, and data between 2048 and 2098 as the \textit{future climate}. The analysis is performed using 8 EURO-CORDEX climate simulations \citep[see][Table 1 for details]{KapicaEtAl2024} forced by the RCP4.5 and RCP8.5 climate scenarios. Results for the more moderate RCP4.5 scenario are presented in Section \ref{sec:res}, while results for the more severe RCP8.5 scenario are provided in the supplementary material (Section 2). 

As \cite{Bloomfield2022} remark, identifying and understanding the impacts of future climate conditions on power systems is a significant challenge, primarily due to the multiple types of uncertainties inherent in Global Climate Models. Results are obtained for the mean of each climate model, though, for concision, we often only present the average results across the climate models. We separately analyse energy droughts in different seasons: winter (DJF), spring (MAM), summer (JJA), and autumn (SON); doing so avoids the droughts clustering in specific seasons with the lowest energy production \citep[see e.g.][]{AllenOtero2023}. While new CMIP6 climate scenarios are available, we restrict attention to CMIP5 projections to permit a comparison with previous studies on future energy droughts \citep{KapicaEtAl2024}.

We separately consider wind and solar production droughts, which correspond respectively to abnormally low wind and solar production. These droughts are defined using the SREPI as described in Section \ref{sec:sei}, with the production $P_t$ equal to the recorded wind or solar capacity factor at time $t$. The total absolute wind (solar) production is equal to the wind (solar) capacity factor multiplied by the installed wind (solar) capacity in the region of interest; hence the capacity factor and production differ by a scaling term. The SREPI is agnostic to the scale of the data, so using capacity factors rather than total production data will not affect the resulting drought characteristics. Ideally, we would also consider droughts defined using the total renewable energy production from both wind and solar power. However, computing the total energy production is not possible without access to the installed wind and solar capacities in each region. The need for more publicly available energy data is discussed further in Section \ref{sec:disc}.

\cite{KapicaEtAl2024} additionally consider the hybrid energy production (HYB), which they define as a linear combination of the normalised wind and solar capacity factors,

\begin{equation}\label{eq:hybrid}
    a \frac{\mathrm{CF}_{SPV,t}}{\mathrm{CF}_{SPV,max}} + (1 - a) \frac{\mathrm{CF}_{WON,t}}{\mathrm{CF}_{WON,max}},
\end{equation}

where $\mathrm{CF}_{SPV,t}$ and $\mathrm{CF}_{WON,t}$ are the capacity factors of the photovoltaic and wind energy generators at time $t$, respectively, $\mathrm{CF}_{SPV,max}$ and $\mathrm{CF}_{WON,max}$ are the maximum value of these capacity factors over a reference period, and $0 \leq a \leq 1$ represents the proportion of solar power in the solar-wind hybrid generator's installed capacity. Due again to the lack of installed capacity data, this proportion is generally not known. We follow \cite{KapicaEtAl2024} by considering this hybrid energy production with $a = 0.5$, implying a 50\% contribution from SPV and 50\% from WON to the energy mix. This choice is discussed further in Section \ref{sec:disc}, and in Section \ref{sec:res}, we present some results that illustrate how the behaviour of energy droughts changes for different values of $a$.

\section{Results}\label{sec:res}

\subsection{The frequency of energy droughts}

\begin{figure}[!htb]
    \centering
    \includegraphics[width=1.0\linewidth]{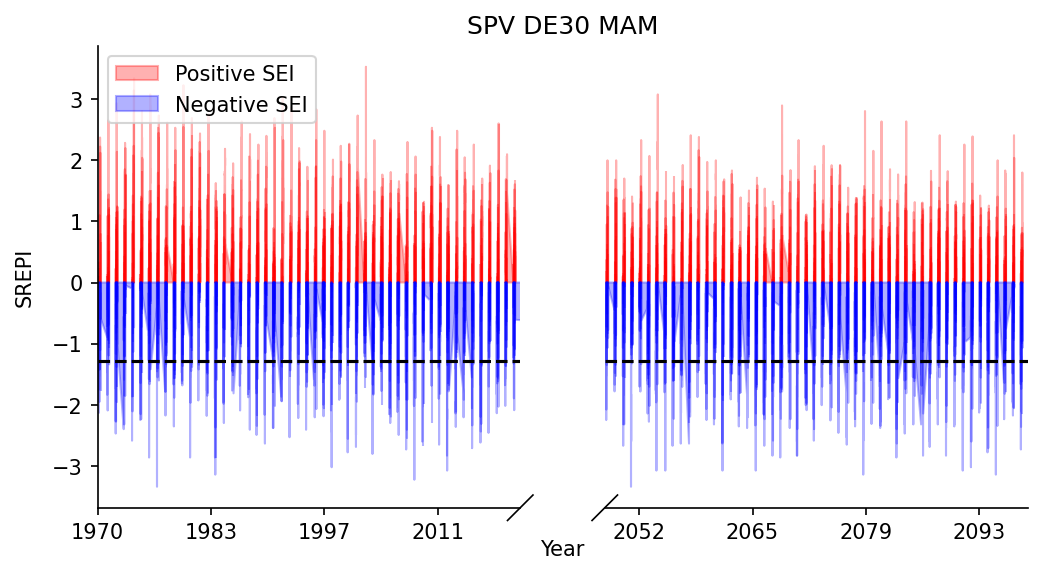}
    \caption{Time series of SREPI for SPV in one region (Berlin, Germany) during spring. Results are shown for the Model 1 mean for historical (1970-2020) and RCP4.5 scenario (2048-2098). A dashed black line is shown at -1.28.}
    \label{fig:SREPI_rcp85_MAM-DE30}    
\end{figure}

In this section, we describe the projected changes in the frequency of energy droughts for SPV, WON, and HYB under the RCP4.5 emissions scenario. Figure \ref{fig:SREPI_rcp85_MAM-DE30} shows the evolution of the SPV SREPI values in spring for the area of Berlin, Germany (the same plot for WON is shown in Figure S5 of the supplementary material). The SREPI behaves similarly between 1970 and 2020, but is generally much lower in the future climate. The SREPI more frequently attains an extremely low value, falling below -1.28 more often than in the present climate, suggesting an increase in solar energy drought frequency and intensity.

\begin{figure}[!htb]
    \centering
    \includegraphics[width=1.0\linewidth]{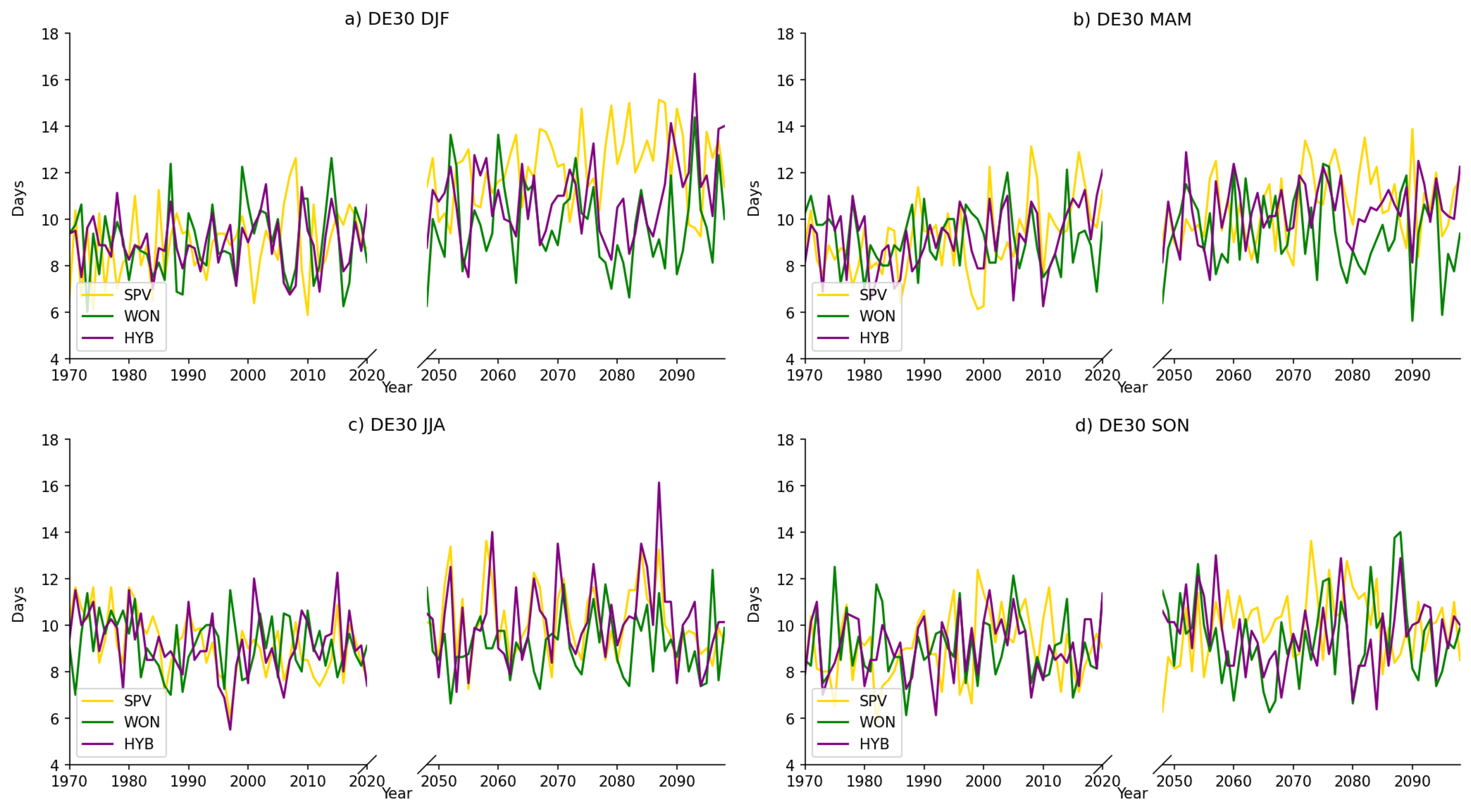}
    \caption{Time series of moderate energy drought days for one region (Berlin, Germany) in different seasons: a) winter, b) spring, c) summer and d) autumn. Multimodel mean for historical (1970-2020) and RCP4.5 scenario (2048-2098).}
    \label{fig:TimeSeries_rcp85_DE11}    
\end{figure}

Figure \ref{fig:TimeSeries_rcp85_DE11} displays time series of annual energy drought days in Berlin, for all seasons and renewable energy sources. Despite some interannual variability, drought days generally increase in the future climate in all seasons, particularly in winter, with solar production droughts increasing more than wind and hybrid production.  The hybrid energy system acts similarly to WON in the colder seasons, and SPV in the warmer seasons. During winter, the total number of solar drought days increases by 37\% from the present to the future climate.

\begin{figure}[!htb]
    \centering
    \includegraphics[width=1.1\linewidth]{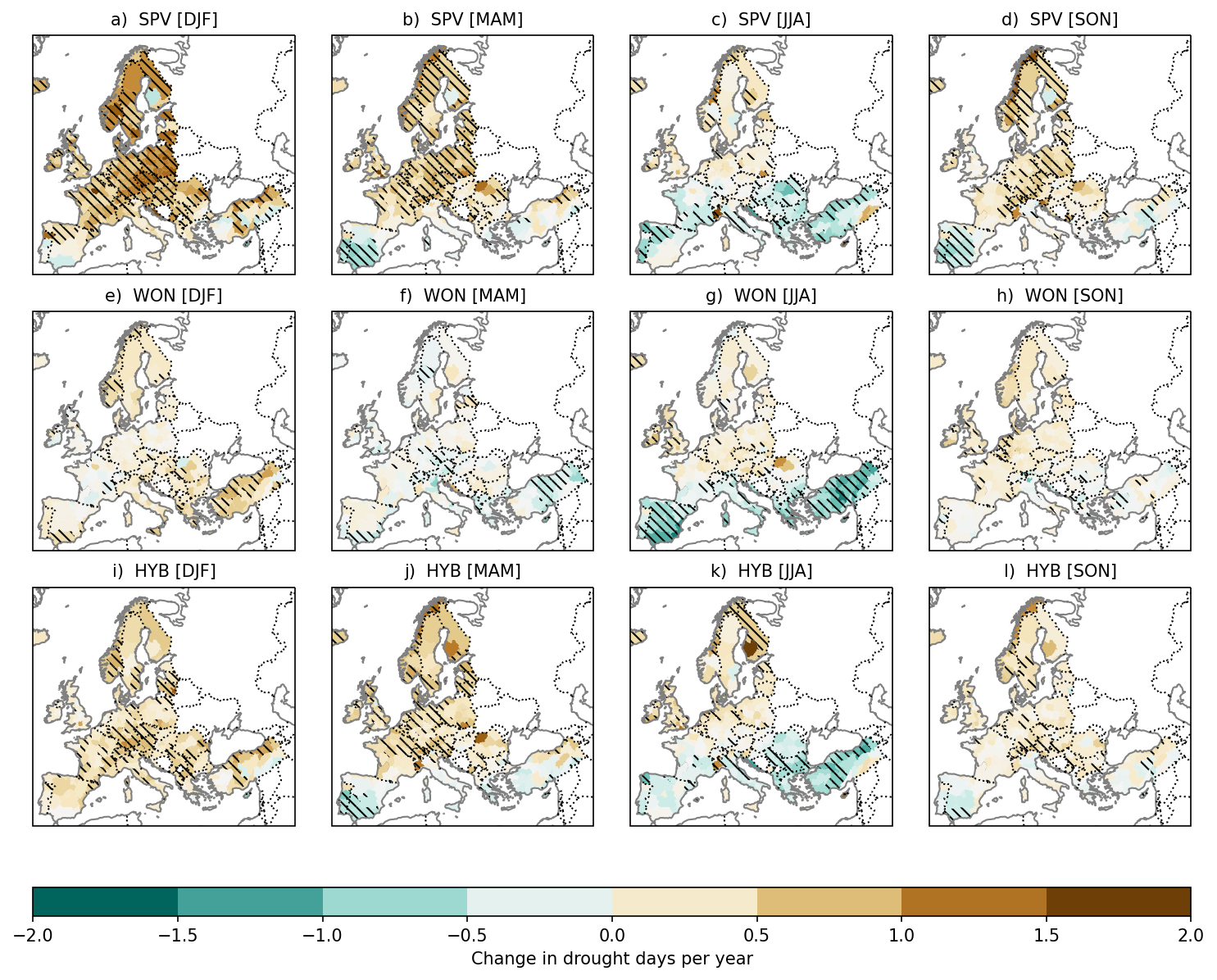}
    \caption{Projected changes in moderate energy drought days per year for SPV (a, b, c, d), WON (e, f, g, h) and Hybrid (i, j, k, l). Each column corresponds to a season: a,e,i) winter, b,f,j) spring, c,g,k) summer and d,h,l) autumn. Multimodel mean for RCP4.5 scenario (2048-2098). Hatches denote regions of ensemble agreement on the sign of bias; i.e., more than 6 of the ensemble members indicate a bias of the same sign.}
    \label{fig:Feq_Energy_drought_Moderate_rcp85}    
\end{figure}

Figure \ref{fig:Feq_Energy_drought_Moderate_rcp85} displays the spatial distribution of these changes in drought frequency. Solar production droughts are expected to become more common in northern and central Europe, particularly in winter and autumn. The number of winter solar drought days is expected to increase by more than 2 per year in Czechia, Poland, Lithuania, Latvia, and Estonia; the colourbar of the plot has been truncated at 2. In the colder seasons, changes of more than 2 days are expected in northern Sweden. Solar production droughts decrease throughout the year in some regions of Turkey and the Iberian Peninsula. A similar pattern is found when using a more extreme threshold to define energy droughts (see Section 3 of the supplementary material).

The projected increase in wind production droughts is markedly lower than for solar. The highest increases in drought frequency occur in Turkey, northern Italy, and western Norway during winter, and in the Scandinavian countries, the UK, and central European countries during summer. In contrast, wind droughts are projected to decrease in summer in southern Europe: wind droughts in Spain, Italy, and Turkey are expected to decrease by more than 2 days per year. Changes are more (less) pronounced in summer (spring). An increase in wind drought frequency over central and northern Europe can be associated with the projected increase in low flow conditions \citep{otero2018assessment}, though the trend in southern Europe is unclear \citep[e.g.][]{CARVALHO201729, ZHANG2018443}.

Changes in HYB droughts resemble the changes to SPV droughts, with larger increases in central and northern Europe. HYB droughts are projected to increase less than SPV droughts in winter and autumn (e.g. in Baltic and Scandinavian countries), but increase more than WON droughts in higher latitudes during summer. In spring, the projected increase of HYB droughts is larger than for WON and SPV droughts in northern Europe. This is consistent with HYB following WON during cold seasons and SPV during warm seasons.

\subsection{The duration and intensity of energy droughts}

\begin{figure}[!htb]
    \centering
    \includegraphics[width=1.1\linewidth]{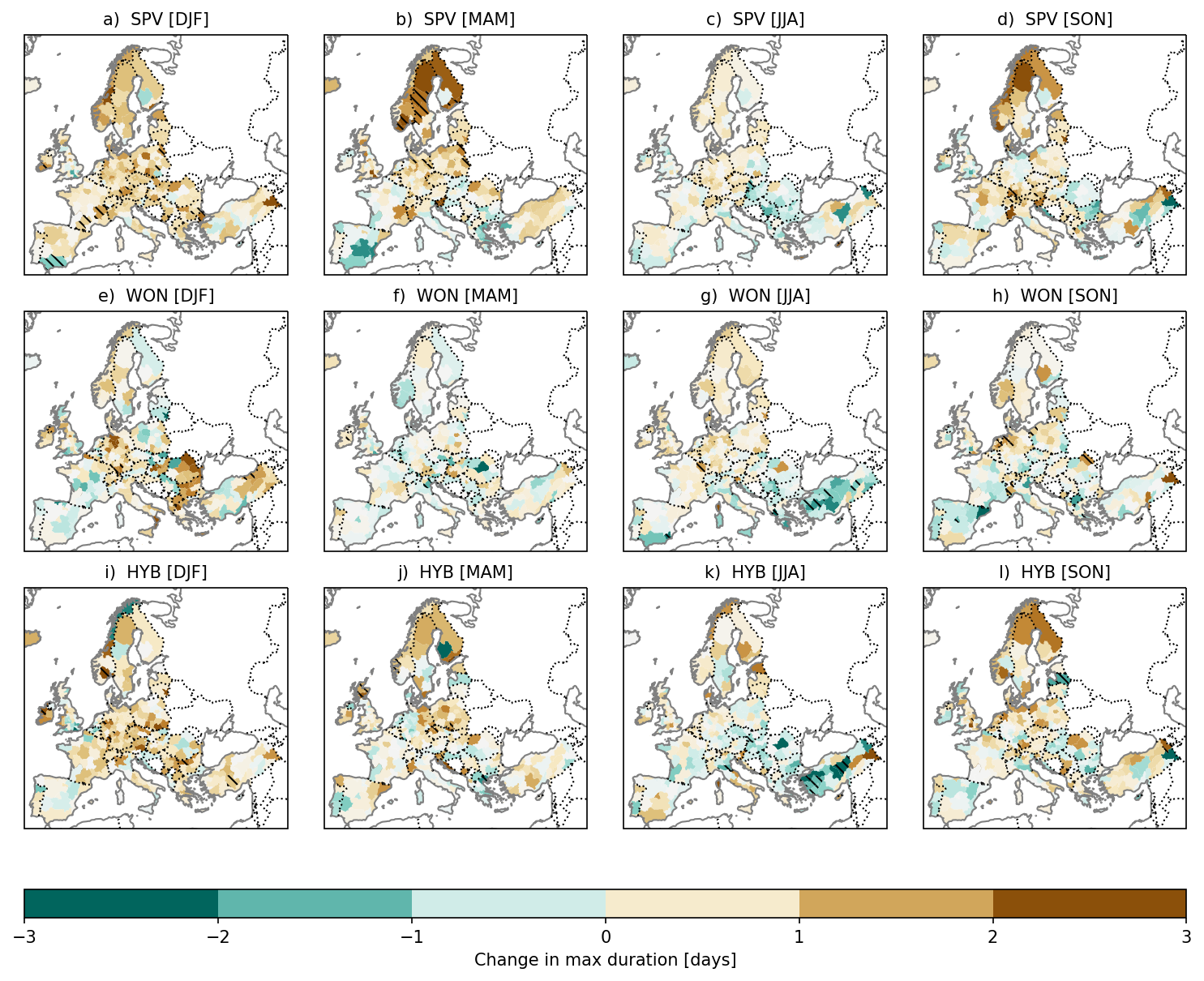}
    \caption{Projected changes in max duration of energy drought days for SPV (a, b, c, d), WON (e, f, g, h) and Hybrid (i, j, k, l). Each column corresponds to a season: a,e,i) winter, b,f,j) spring, c,g,k) summer and d,h,i) autumn. Multimodel mean for RCP4.5 scenario (2048-2098). Hatches denote regions of ensemble agreement on the sign of bias; i.e., more than 6 of the ensemble members indicate a bias of the same sign.}
    \label{fig:Energy_drought_rcp85_dur_Max}    
\end{figure}

In the following, we analyse changes in energy drought duration and intensity. Figure \ref{fig:Energy_drought_rcp85_dur_Max} presents changes in the maximum drought duration in each region. The spatial variability of the maximum drought duration is larger than for the mean duration (see supplementary material, Figure S11), and the climate models only agree on the sign of the change in duration in a few regions. These results should therefore be interpreted with caution. For SPV droughts, an increase in the maximum drought duration is clear during winter, spring, and autumn in northern Europe: increases of more than 2 days are projected in Scandinavian countries.

Changes in maximum wind drought duration are less clear. We find a more pronounced spatial pattern during summer and autumn, where more persistent wind droughts are projected in Scandinavia, and less persistent wind droughts in the Iberian Peninsula and Turkey; this is similar to the mean drought duration. Similarly, the spatial pattern for the maximum duration of HYB droughts is complex, and few models agree on the changes. The most evident changes are more persistent HYB droughts in northern Europe, with an increased duration of more than 2 days.

These results align with those in \cite{KapicaEtAl2024}, highlighting that droughts defined using SREPI values exhibit similar properties to those defined using fixed production thresholds. One benefit of employing standardised indices is that we can additionally study the magnitude and intensity of energy droughts across different regions. The magnitude of an energy drought takes into account both its duration and intensity, and changes in drought magnitude are therefore similar to changes in the mean duration (see Figure S11 of the supplementary material). Figure \ref{fig:Energy_drought_rcp85_dur_sev} displays the projected changes in average drought intensity at each region; recall that the average drought intensity is the average SREPI value over the drought event. The intensity is defined on the same scale as the standardised indices. Solar production drought intensity increases in almost all regions and seasons, except in summer, when a decrease in solar drought magnitude is projected in the Iberian Peninsula and some regions in southeast Europe. The HYB energy droughts follow a similar pattern, whereas changes in the average intensity of wind production droughts tend to behave similarly to changes in mean occurrence. For instance, a decrease in intensity is projected at lower latitudes and an increase at higher latitudes; this dipole pattern is more evident in summer. Following \cite{BrackenEtAl2023}, we computed the Compound Drought Magnitude, which accounts for simultaneous solar and wind droughts. These results are comparable to the mean intensity of the hybrid generator (Fig. \ref{fig:Energy_drought_rcp85_dur_sev}i,j,k,l), although they exhibit greater spatial variability and lower model agreement (see Fig. S4 for the historical period and Fig. S13 for the RCP4.5 scenario).

Changes in energy drought frequency, duration, and intensity are largest for solar production, particularly over Scandinavia. Since, solar power production is largely determined by global horizontal irradiation and 2m temperature, we explore how these results correspond to projected changes in total cloud cover (supplementary material, Figure S14). There is a clear increase in cloud cover over Scandinavia in all seasons, while a decrease in cloud cover is projected over the Iberian Peninsula for spring, summer, and autumn; this is a region with less increase in SPV energy droughts. This supports the conjecture from \cite{KapicaEtAl2024} that a decrease in solar power production can be attributed to more storms and their associated clouds, which decreases solar radiation and therefore increases the likelihood of a solar production drought.

\begin{figure}[!htb]
    \centering
    \includegraphics[width=1.1\linewidth]{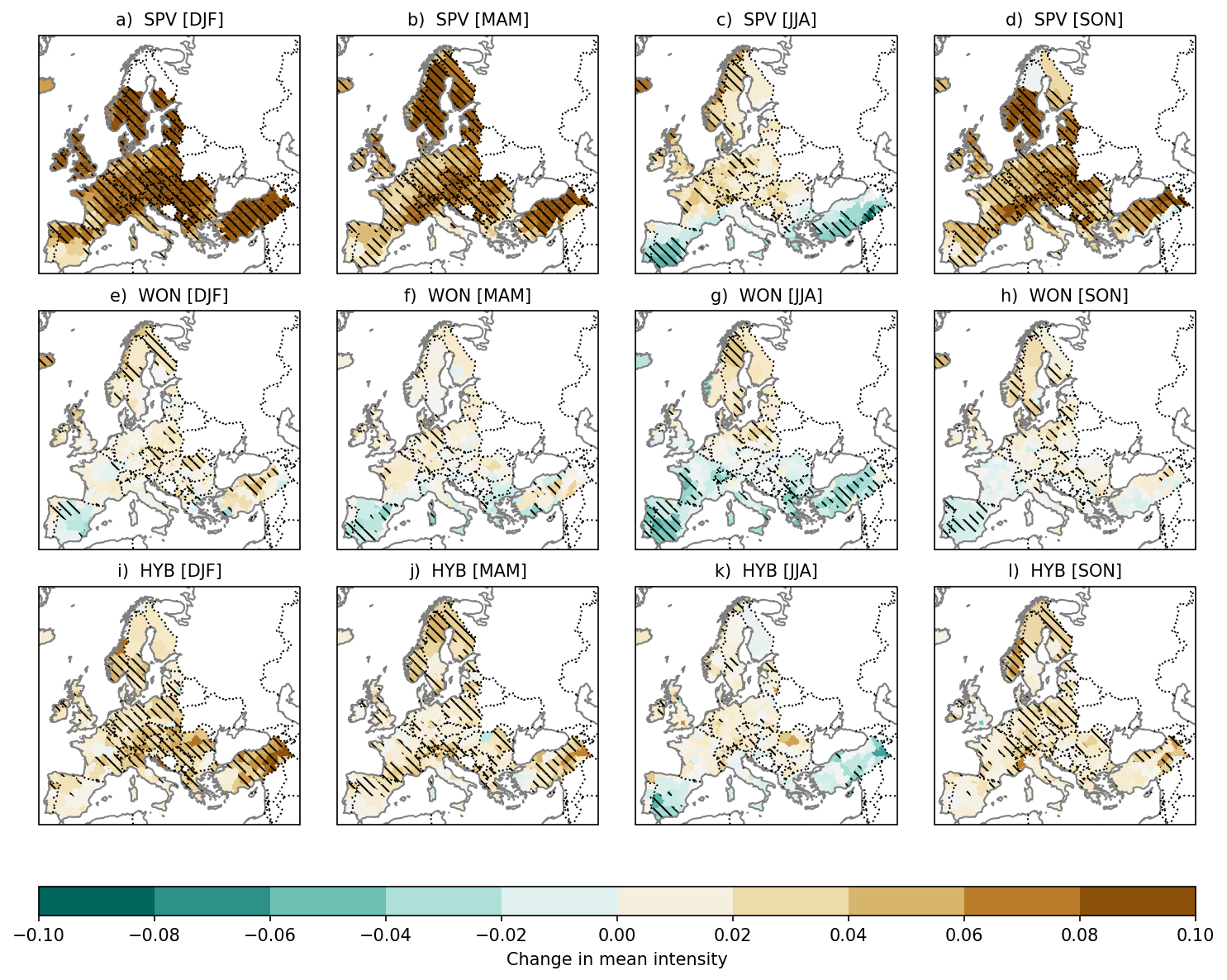}
    \caption{Projected changes in average drought intensity for SPV (a, b, c, d), WON (e, f, g, h) and HYB (i, j, k, l). Each column corresponds to a season: a,e,i) winter, b,f,j) spring, c,g,k) summer and d,h,i) autumn. The average intensity is the average SREPI value across the drought event. Multimodel mean for RCP4.5 scenario (2048-2098). Hatches denote regions of ensemble agreement on the sign of bias; i.e., more than 6 of the ensemble members indicate a bias of the same sign.}
    \label{fig:Energy_drought_rcp85_dur_sev}    
\end{figure}

\subsection{Energy system adaptation}

The previous section illustrated that solar power production droughts are projected to become more common, persistent, and intense as a result of climate change, particularly in central and northern Europe. In this section, we explore two possible ways to mitigate the changes in energy droughts: increasing the installed capacity, and varying the contribution of solar and wind ($a$) to the hybrid solar-wind generator.

Increasing the installed wind and solar capacities allows us to study how much additional installed capacity is required for energy droughts to not become more common or intense in the future. This will be important for policymakers when deciding how to configure energy systems in the future. We multiply the capacity factors in the future climate by a factor of 1.1 (10\% increase), 1.25 (25\% increase), and 1.5 (50\% increase). The present-day time series remains unaltered, since we want to study how changes in the energy drought system can affect the resulting drought characteristics relative to the current energy system. 

Changes in the frequency of future wind droughts for the different hypothetical increases of installed capacities are shown in supplementary material (Figs. S15 and S16). Results are shown as a percentage decrease relative to when the installed capacity is unaltered. Unsurprisingly, increasing the installed capacity severely decreases the frequency of future energy droughts, particularly in southern Europe during summer. Similar changes are found for SPV, with a smaller reduction in winter and autumn in northern Europe (see supplementary material, Figure. S16).

Previous studies have also considered changing the relative contribution of wind and solar to the energy mix, finding that a more balanced mix can help to mitigate the effects of weather events that cause low wind \emph{or} low solar power production. We therefore study how changing the factor $a$ in the hybrid solar-wind generator in Equation \ref{eq:hybrid}. We compute the changes in energy drought frequency with a solar fraction ($a$) equal to 0.2 (20\% solar energy and 80\% wind energy) and 0.8 (80\% solar energy and 20\% wind energy).

\begin{figure}[!htb]
    \centering
    \includegraphics[width=1.1\linewidth]{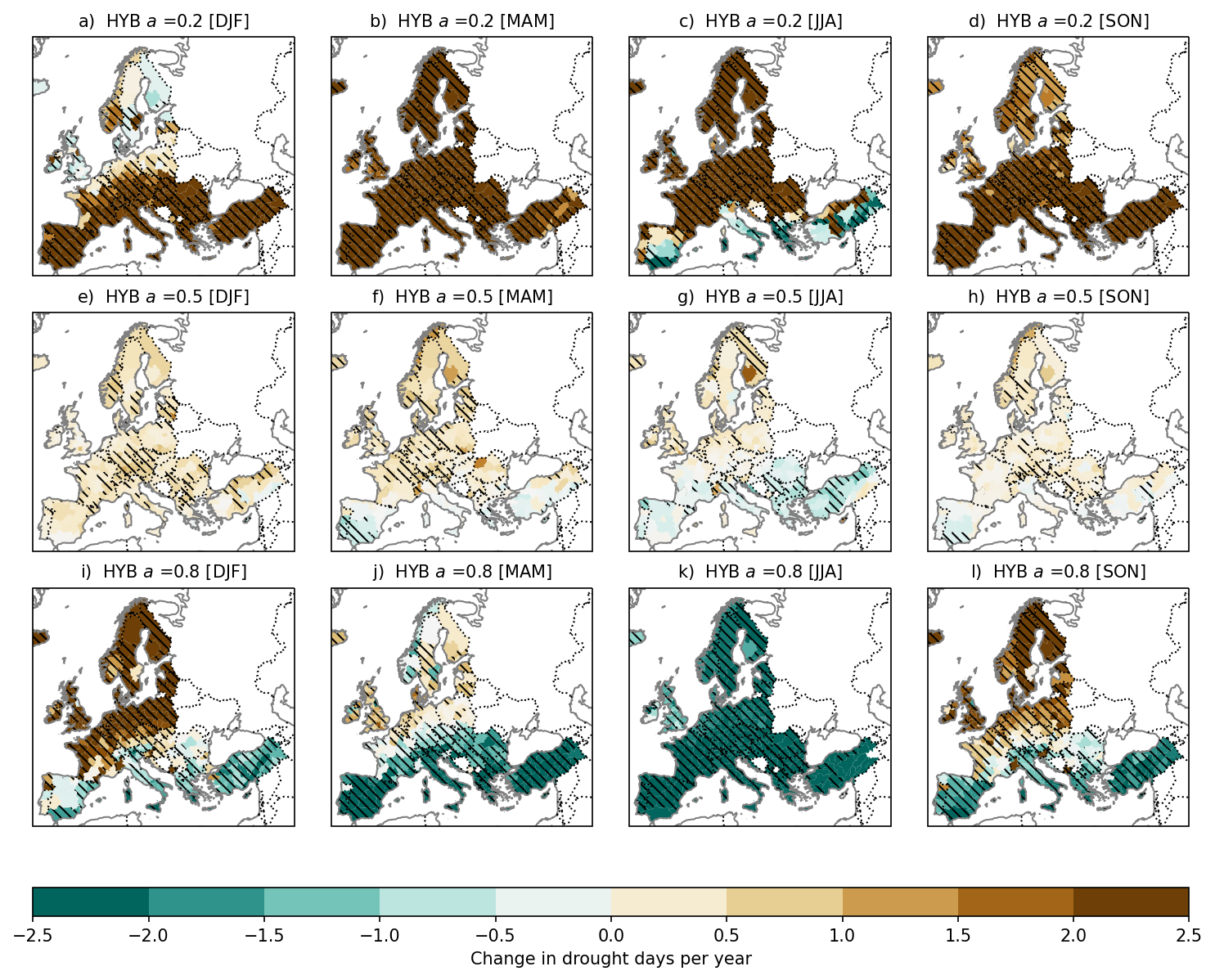}
    \caption{Projected changes in moderate energy drought days for  Hybrid generation. CMIP5 multimodel mean for RCP4.5 scenario (2048-2098). Future shared fraction ($a$) of (a,b,c,d) 20\%, (e,f,g,h) 50\% and (i,j,k,l) 80\%. Each column corresponds to a season: a,e,i) winter, b,f,j) spring, c,g,k) summer and d,h,i) autumn. Hatches denote regions of ensemble agreement on the sign of bias; i.e., more than 6 of the ensemble members indicate a bias of the same sign.}    \label{fig:Feq_Energy_drought_Moderate_rcp85_HYB_ca_altered}    
\end{figure}

Figure \ref{fig:Feq_Energy_drought_Moderate_rcp85_HYB_ca_altered} shows the projected changes in HYB droughts in the future, relative to the hybrid scenario in the present climate. If the solar contribution is decreased to 20\%, then the number of energy droughts is projected to increases in almost all regions. If the solar contribution is increased to 80\%, then energy production droughts will become less frequent in southern Europe, but more common in northern Europe during winter, spring, and autumn. This aligns with previous results regarding the occurrence of cloud cover: more storm tracks and clouds in Scandinavia mean there is no benefit to increasing the installed solar capacity in these seasons. However, due to the increased solar radiation caused by climate change, it becomes beneficial to increase installed solar capacity in southern Europe. There is comparatively less benefit to increasing the proportion of installed wind capacities.

\section{Summary}\label{sec:conc}

In this study, we analyse projected changes in the frequency, duration, and intensity of wind and solar production droughts using standardised energy indices, and investigate how potentially impactful changes can be mitigated by increasing the installed capacity or adapting the mix of renewable energy sources in future energy systems. Our results support previous conclusions regarding the frequency and duration of energy droughts under future climate projections, but by employing the SREPI to define energy droughts, we can additionally compare changes to the magnitude and intensity of energy droughts in different regions. 

Using EURO-CORDEX climate simulations (based on CMIP5), we find that solar production will be more heavily affected by changes to the climate than wind production, particularly at high latitudes. This results in solar production droughts in Scandinavia that occur more frequently, persist for longer, and are more intense. We illustrate that this coincides with a projected increase in total cloud cover over this region, which aligns with previous research on the behaviour of storm tracks and extratropical cyclones in Europe. In contrast, the increased solar radiation caused by climate change will result in greater solar power production in southern European countries, decreasing the impacts associated with energy droughts.

This motivates future studies on the interaction between weather patterns and future energy droughts. Extratropical cyclones are projected to shift eastward in the North Atlantic basins, resulting in increased cloud clover over northern Europe \citep{zappa2013multimodel,dolores2022future,priestley2022cyclones}, but energy droughts are also affected by atmospheric blocking, which does not exhibit clear patterns in future climate scenarios \citep{woollings2018blocking,davini_cmip3_2020}. Future work into the effects of atmospheric blocking on energy droughts in future climate projections would therefore help to provide additional meteorological reasoning for the results seen herein.In particular, it would help to attribute changes in wind power production, which is affected by multiple convoluted factors.  Moreover, additional efforts should focus on quantifying the impacts of atmospheric blocking on energy demand under climate change, which is greatly influenced by weather patterns \citep{Otero2022}. This could additionally be studied in the context of compound drought events (co-occurrence of wind and solar droughts), which have the potential to generate particularly large impacts on energy systems.

To counteract increases in frequency and intensity of energy production droughts, we study two different strategies to adapt the energy mix: increasing the installed capacity, and varying the ratio of wind and solar in a hybrid energy generator. The SREPI allows droughts to be defined in terms of the present climate, allowing us to compare results to the situation where no action is taken to mitigate climate change. Unsurprisingly, increasing the installed capacity drastically decreases the frequency, duration, and intensity of drought events. It is particularly beneficial to increase solar capacity in southern Europe, to exploit the increased solar radiation. Due to the presence of storms and clouds in northern Europe, increasing wind capacity reduces the frequency of energy droughts in winter and autumn in this region, though in summer and spring it is generally still more beneficial to increase solar capacity.

\section{Discussion: The need for more energy data}\label{sec:disc}


The goal of this analysis is to provide information that may be useful for policymakers to understand what action, if any, is required to counteract the effects of climate change on renewable energy systems. However, for such studies to be more directly relevant for policymaking, we argue that more publicly available energy data is needed. We therefore end this paper with a discussion of the limitations of this study that arise due to a lack of data, and advocate further work to make such data available.

The study herein does not consider off-shore data, despite the large number of wind and solar power plants based off-shore, and only considers wind and solar power, despite other sources of renewable energy contributing to national energy mixes, most notably hydropower. We have limited information regarding the physics and management of power systems, including the efficiency of renewable plants when harvesting wind and solar energy, the plants' operational hours, and their ability to store energy. Similarly, we make the simplifying assumption that higher wind speeds lead to increased wind power production, disregarding that wind turbines often cannot operate under high wind speeds. Additional data is required to incorporate these aspects into the analysis, and it would therefore be incredibly beneficial to have a unified data set that contains all of this information.


Furthermore, our results are specific for each region considered in the study. That is, we assume that the risks of a power outage are determined uniquely by the energy production at that region. Such a framework allows us to analyse local characteristics of the energy system, acknowledging that the climate, and hence renewable energy production, will evolve differently at a sub-country level. However, energy systems are typically operated at a national level, with energy sharing between countries also possible. Having a reasonable representation of how renewable energy resources are shared between regions and countries would permit a more realistic assessment of the risks of energy droughts. Future studies could then investigate how energy sharing strategies could be adapted to reduce the risks of energy droughts.


We restrict attention throughout to renewable energy production droughts based on the SREPI. However, \cite{vanderWiel2019} additionally study energy supply droughts that depend on the residual load (demand minus production), which can similarly be defined in terms of the Standardised Residual Load Index \citep[SRLI;][]{AllenOtero2023}. Since energy demand is itself highly weather dependent, it would be informative to study projections of the SRLI in future climate scenarios, which would allow us to analyse how energy supply droughts are expected to evolve in the future. However, such an analysis is currently not possible, since we do not have access to up-to-date installed capacity data in each region of interest, and therefore cannot calculate the total renewable energy production. 


Future work would therefore greatly benefit from having up-to-date installed wind and solar capacities available, at either a regional or national level. The hybrid generator defined in Equation \ref{eq:hybrid} provides a simple approach to simulate the total renewable energy production \citep{KapicaEtAl2024}, but the assumption that wind and solar each contribute 50\% to the renewable energy mix (in all regions) is somewhat arbitrary, and is generally not representative of the operational set up in the regions considered. However, this provides a necessary and useful baseline in the absence of specific regional capacity data.


Finally, an analysis of energy supply droughts would also require an accurate estimate of energy demand in the future. Energy demand is typically estimated from weather data via the number of hot and dry days, though this additionally introduces uncertainty that should be accounted for. In particular, such an approach neglects other important predictors of energy demand, such as the prevailing socio-economic conditions. However, designing a more complex model for energy demand is hindered by the lack of prolonged, historical time series of regional energy demand. Similarly, there are also uncertainties in the future climate simulations, as well as the corresponding energy indicators from the \cite{cs32021}. While these have been bias-corrected, some biases in the EURO-CORDEX data may remain; \cite{Bartok2017projected} find that solar irradiance is generally over-estimated, for example.

\ack
We are very grateful to Hannah Bloomfield and Jacek Kapica for useful comments during the preparation of this work. 
EDT acknowledgments the nextGEMS project.

\bibliography{references}

\end{document}